\def\spose#1{\hbox to 0pt{#1\hss}} %
\def\Dt{\spose{\raise 1.5ex\hbox{\hskip3pt$\mathchar"201$}}}    
\def\dt{\spose{\raise 1.0ex\hbox{\hskip2pt$\mathchar"201$}}}    

\def\dotdeg{\hbox{$^\circ$\hskip-3pt .}}

\def\lta{\mathrel{\spose{\lower 3pt\hbox{$\mathchar"218$}}
     \raise 2.0pt\hbox{$\mathchar"13C$}}}

\def\gta{\mathrel{\spose{\lower 3pt\hbox{$\mathchar"218$}}
     \raise 2.0pt\hbox{$\mathchar"13E$}}}

\documentstyle[12pt,psfig]{article}
\begin{document}

\begin{center}
				Radial Velocities of Stars

				with Multiple Co-orbital Planets

\vspace*{0.5in}
				Anthony R. Dobrovolskis 

					SETI Institute

				245-3 NASA Ames Research Center 

				Moffett Field, CA 94035-1000 

\vspace*{0.5in}
					2014 April 21

\vspace*{0.5in}
					ABSTRACT
\end{center}

To date, well over a thousand planets have been discovered 
orbiting other stars, hundreds of them in multi-planet systems. 
Most of these exoplanets have been detected by either 
the transit method or the radial velocity method, 
rather than by other methods such as astrometry or direct imaging. 
Both the radial velocity and astrometric methods 
rely upon the reflex motion of the parent star 
induced by the gravitational attraction of its planets. 
However, this reflex motion is subject to misinterpretation 
when a star has two or more planets with the same orbital period. 
Such co-orbital planets may effectively ``hide'' from detection 
by current algorithms. 

In principle, any number of planets can share the same orbit; 
the case where they all have the same mass has been studied most. 
Salo and Yoder (A \& A 205, 309--327, 1988) have shown that 
more than 8 planets of equal mass sharing a circular orbit 
must be equally spaced for dynamical stability, 
while fewer than 7 equal-mass planets are stable 
only in a configuration where all of the planets 
remain on the same side of their parent star. 
For 7 or 8 equal-mass planets, both configurations are stable. 

By symmetry, it is clear that the equally-spaced systems produce 
no reflex motion or radial velocity signal at all in their parent stars. 
This could lead to their being overlooked entirely, 
unless they happen to be detected by the transit method. 
It is equally clear that the lopsided systems produce 
a greater radial velocity signal than a single such planet would, 
but a smaller signal than if all of the planets were combined into one. 
This could seriously mislead estimates of exoplanet masses and densities.
Transit data and ellipsoidal (tidal) brightness variations in such systems 
also are subject to misinterpretation. 
This behavior is also representative of more natural systems, 
with co-orbital planets of different masses. 

\newpage

\section{INTRODUCTION}

It has long been known that three objects (stars, planets, satellites, 
or some combination), all circling their mutual center of mass 
at the vertices of an equilateral triangle, are in a state of equilibrium, 
independent of their individual mass values $m_a$, $m_b$, $m_c$ (Lagrange, 1772). 
Furthermore, this equilibrium is stable, provided that 
\begin{equation}
		m_a m_b + m_a m_c + m_b m_c < M^2/27 ,
\end{equation}
where $M \equiv m_a+m_b+m_c$ is the total mass of the system (Routh 1875). 
The latter criterion requires that one body be $\gta$ 26 times more massive than 
the other two combined (see Dobrovolskis, 2013, hereinafter referred to as Paper 1). 

Such configurations are known as Trojan systems, and their effects on the 
reflex motion of the parent star are well understood; see Paper 1 for a review. 
It is less well known that any number $N > 1$ of secondaries can be in 
equilibrium while sharing the same circular orbit around their primary; 
the case where they all have the same mass has been studied most. 
For small $N$, there are numerous equilibrium configurations, 
some stable, and some unstable. All such configurations affect 
the the radial velocity variations and reflex motion of the parent star, 
and some produce no reflex motion at all! 

This paper examines the effects of such co-orbital systems of planets 
on the reflex motions of their primary stars, and also on tides and transits. 
These effects are significant because they affect the interpretation 
of radial velocity and astrometric data. Most techniques currently used to 
analyze such data assume that no two planets share the same orbital period. 
Thus exoplanets with the same period can effectively ``hide'' from detection. 

\section{SALO SYSTEMS}

Salo \& Yoder (1988) undertook a numerical search for the stationary 
configurations of systems of multiple planets (or satellites) 
sharing the same circular orbit about a central star (or planet),  
and they performed ``A reasonably thorough examination of the coplanar, 
nearly circular particle rings with $N \le 9$'' (Salo \& Yoder, 1988). 
Their study was quite thorough indeed, 
and much of the following analysis is based on their paper, 
hereinafter referred to as ``SY88''. 

Like SY88, I too use $N$ to denote the number of orbiting bodies, 
or secondaries, $m_i$ for their individual masses, 
and $r_i$ for their distances from the central body, or primary. 
However, note that SY88 used $\gamma$ for the Newtonian constant 
of gravitation, and $M$ for the mass of the primary; 
but I use $G$ for the gravitational constant, 
and $m_a$ for the greatest mass, reserving $M$ for 
the total mass of the system: $M \equiv m_a +\sum m_i$. 
Note also that SY88 did not distinguish between the total mass $M$ and the 
mass $m_a$ of the primary, developing their theory only to order $m_i/m_a$. 
(Henceforth I adopt the common usage $\mu_i \equiv m_i/m_a$.) 
SY88 included both direct and indirect perturbations 
between the secondaries, though only to order $\mu_i$. 

\newpage

\subsection{Equilibria}

It is not guaranteed that dynamical equilibria even exist in this problem 
(except for the equilateral triangle configuration for $N$=2), much less 
that they are unique. In fact, unlike the equilateral solution for $N$=2, 
the equilibrium configurations for $N \ge 3$ generally depend on the 
individual masses $m_i$ (SY88, Renner \& Sicardy 2004). 
Because the corresponding parameter space is so vast, 
SY88 restricted their search space for $N > 3$ 
to all $m_i$ equal, so that $M = m_a +N m_i$. 
For simplicity, I refer to such arrangements as ``Salo systems''. 

2.1.1: Families

Even under this constraint of equal-mass secondaries, 
SY88 found at least one equilibrium for each value of $N$, 
and sometimes several; as they wrote, 
``we initially had no insight that the number and complexity 
of the stationary configurations would be so rich.'' 
SY88 found that the equilibria fall into three families, 
plus three variant solutions (see Fig. 2 of SY88): 

The simplest family is the one SY88 twice called ``trivial'' 
and chose to label as ``Type II''. 
By symmetry, this configuration exists for all $N > 1$, 
and consists of $N$ equal-mass secondaries, 
all at strictly identical distances from the primary 
and equally spaced about it. 
This family was known to Maxwell (1859). 

In contrast, the family SY88 twice called ``compact'' and labeled ``Type Ia'' 
exists only for $2 \le N \le 8$, and consists of lop-sided configurations 
where all of the $N$ secondaries are concentrated 
more or less on the same side of the primary. 
The equilateral triangle configuration with $N$=2 
and $m_i=m_b=m_c$ is the first member of family Ia. 

The family SY88 called ``Type IIIa'' exists only for $3 \le N \le 7$, 
and consists of configurations where $N-1$ secondaries 
are concentrated on the opposite side from the remaining one. 

Finally, SY88 also found two additional solutions for $N$=7, 
which they labeled ``Ib'' and ``IIIb'', and one more for $N$=8, 
which they called ``Ib'' as well. They considered these 
as variants of the corresponding Type Ia and IIIa solutions, 
but with slightly wider separations between secondaries. 
However, as discussed below, 
I regard solution ``Ib'' for $N$=7 as variant ``IIb'' instead, 
and all those of the equally-spaced family ``II'' as ``Type IIa''. 

Table 1 of SY88 lists the angular locations of each secondary 
in all of the equilibria for $2 \le N \le 9$. Some of their 
tabulated angular positions for Type IIa are off by $\sim0\dotdeg001$, 
so I presume that their other locations have comparable accuracy. 
To facilitate comparison among the solutions, 
I have subtracted $180^\circ$ from their tabulated positions 
for all solutions of Types Ia and Ib, 
but not for those of Types IIa, IIb, IIIa, or IIIb. 

The figure resembling a dartboard (Fig. 1) plots 
the resulting angular locations $\theta_i$ of the secondaries 
for all of the equilibria of Types Ia, Ib, IIIa, and IIIb, 
as well as those of Types IIa and IIb for $N$=7. 
Note that this plot is symmetric about the vertical axis through 
the center, and the origin of longitude ($\theta$=0) is at the top. 
The resulting figure has the same orientation as Fig. 3 of SY88; 
but compared to Fig. 2 of SY88, 
all solutions of Types Ia and Ib are rotated $90^\circ$ clockwise, 
while all those of Types IIa, IIb, IIIa, and IIIb 
are rotated $90^\circ$ counter-clockwise, for consistency. 

\begin{figure}
\centerline{\psfig{figure=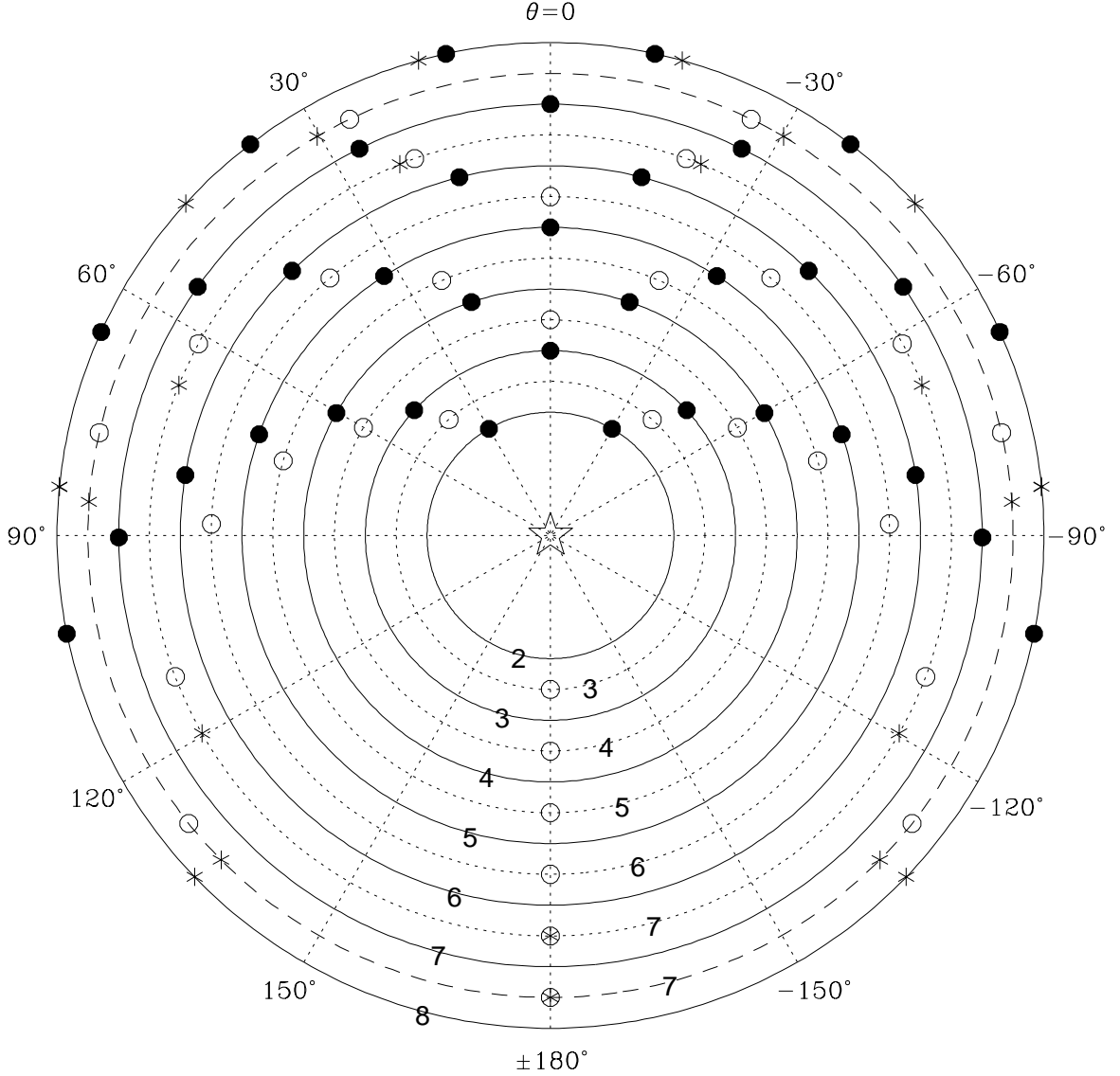,width=6.5in,height=6.5in}}
\caption{ Equilibrium angular locations of $N$ secondaries of equal mass 
sharing the same circular orbit about a much more massive primary. 
From innermost to outermost, the large solid circles represent the Type I orbits 
for $N=2$ to $N=8$, as labeled; the solid dots ($\bullet$) show the locations 
of the secondaries for sub-type Ia (stable), while the asterisks ($\ast$) 
show the locations of the secondaries for sub-type Ib ($N=8$ only, unstable). 
Also from innermost to outermost, the large dotted circles represent the 
Type III orbits for $N=3$ to $N=7$, as labeled; the open dots ($\circ$) show 
the locations of the secondaries for sub-type IIIa (unstable), while the asterisks 
show the locations of the secondaries for sub-type IIIb ($N=7$ only, also unstable). 
The large dashed circle represents the Type II orbits for $N=7$ only; 
the open dots show the locations of the secondaries for sub-type IIa (stable), 
while the asterisks show the locations of the secondaries 
for sub-type IIb (unstable). }
\end{figure}

Comparing adjacent solid and dotted circles (orbits) in Fig. 1 
then shows that the Type IIIa configuration for a given $N$ 
derives from the Type Ia solution for $N-1$ 
by adding another secondary of the same mass at $\theta=180^\circ$, 
and slightly spreading the other secondaries along their orbit 
(with concomitant reduction in $A$; see below). 
This relationship probably could be clarified by reducing the mass 
of the ``new'' secondary at $180^\circ$ gradually back to zero. 

Except for $N$=7, Fig. 1 does not display the solutions of Type II. 
However, when these are all oriented so that one secondary always resides at 
$\theta=180^\circ$, a similar comparison suggests that the Type IIa solution 
for a given $N$ also is related to the Type IIIa solution for the same $N$. 
Comparing the dashed circle in Fig. 1 with the outermost dotted circle 
supports this relationship for $N$=7. 

Furthermore, Fig. 1 reveals that the equibrium solution for $N$=7 which 
SY88 call ``Type Ib'' is not really related closely to solution Ib for $N$=8, 
nor to any member of family Ia either. 
Instead, the dashed circle shows that this equilibrium is best regarded as 
a non-trivial variant of the equally-spaced Type II configuration for $N$=7. 
Therefore I refer to this slightly asymmetric solution as Type IIb, 
and to the strictly equally-spaced family as Type IIa, as mentioned above. 

In this revised classification, 
each family has one ``black sheep'' (its variant solution b). 
Finite families I and III branch into two variants (a and b) 
only when they terminate (at $N$=8 and $N$=7 respectively), 
while infinite family II bifurcates only where it crosses 
its stability threshold at $N$=7 (see below). 

2.1.2: Stability 

Even when dynamical equilibria do exist, 
it is not guaranteed that any are stable. 
In order to determine their stability, 
for each equilibrium solution with $N$ secondaries, 
SY88 found spectra of $N$ real eigenvalues $\lambda$. 
For family IIa, all but one or two of these eigenvalues 
come in identical pairs; while for all families, 
exactly one of these eigenvalues is always zero, 
corresponding to an arbitrary rotation of the whole system, 
and is disregarded. To lowest order in the mass ratio $\mu_i$, 
each positive eigenvalue corresponds to a mode of infinitesimal 
harmonic oscillation with a period of $ P_0 / \sqrt{ \lambda \mu_i } $, 
while each negative eigenvalue corresponds to 
an exponentially growing unstable mode with an $e$-folding 
time of $ P_0 / ( 2\pi \sqrt{ -\lambda \mu_i } ) $. 
A given configuration is unstable as a whole 
if any of its eigenvalues are negative. 

Fig. 4 of SY88 plots their eigenvalue spectra for families Ia, 
IIa, and IIIa (note that the Type IIa spectrum for $N$=12 
is missing one dot at $\lambda$ = 175.379; H. Salo, personal 
communication). Judging by the spectrum they plotted for Type III, 
$N$=7, the eigenvalues listed in Table 1 of SY88 for the Type IIIa 
and Type IIIb solutions appear to have been switched; 
but their tabulated angular positions seem to be correct, 
from comparison with Fig. 2 of SY88. Because the gravitational 
potential energies listed in Table 1 of SY88 differ by $\lta$ 0.003 \% 
among configurations IIIa, IIIb, IIa, and IIb=``Ib'' for $N$=7, 
I cannot tell whether these have been switched as well. 

My Table 1 lists the algebraically least non-zero eigenvalue $\Lambda$ 
(re-arranged as above from SY88) for each of the equilibria 
with $2 \le N \le 9$, along with certain other parameters of interest. 
Based on the sign of $\Lambda$, 
SY88 found that all Type Ia equilibria are stable, 
but that Type IIa solutions are stable only for $N \ge 7$, 
while all of the other solutions are unstable. 
SY88 confirmed this transition for Type IIa equilibria 
by numerical integrations, as displayed in their Fig. 11; 
but note that they mistakenly referred to these solutions 
as ``Type III'' in their explanatory text. 

\begin{tabular}{|c||c|c|c|c|c|c|c|c|c|}
\hline
  Type  &	$N$	&	2	&	3	&       4       &	5	&       6       &       7       &       8       &   9   \\
\hline
\hline
	&	$\Lambda$&	13.5000	&	14.9292	&	15.1258	&	14.2333	&	12.6011	&	9.9030	&	5.8509	&	\\
	&	$A$	&	.86603	&	.78492	&       .72234  &	.66790	&       .61619  &       .56231  &       .49621  &	\\
   Ia	&	$T$	&	.50000	&	.27845	&       .14478  &	.05301	&       -.01431 &       -.06510 &       -.10231 &	\\
        &	$\Delta$&	--	&	--	&       --      &	--	&       --      &$180\dotdeg552$&       --      &	\\
        &       $\delta$& $60\dotdeg000$& $47\dotdeg361$& $37\dotdeg356$& $32\dotdeg660$& $28\dotdeg536$& $26\dotdeg278$& $24\dotdeg460$&     
  \\
        &       $\mu_H$ &       .01200  &       .00622  &       .00315  &       .00213  &       .00144  &       .00113  &       .00091  &     
  \\
\hline
	&	$\Lambda$&		&       	&               &       	&               &               &	-2.0120	&       \\
        &	$A$	&       	&       	&               &       	&               &               &       .26065  &	\\
   Ib	&	$T$	&       	&       	&               &       	&               &               &       -.06415 &	\\
	&	$\Delta$&      		&       	&               &       	&               &               &$181\dotdeg522$&	\\
        &	"	&       	&       	&               &       	&               &               &       "       &	\\
\hline
\hline
	&	$\Lambda$&	-5.2500	&	-2.3349	&	-2.0680	&	-1.5500	&	-0.8349	&	0.0426	&	1.0587	& 2.1955  \\
        &	$A$	&	0	&	0	&       0       &	0	&       0       &       0       &       0       & 0	  \\
  IIa   &	$T$	&	1.00000	&	0	&       0       &	0	&       0       &       0       &       0       & 0	  \\
	&	$\kappa$&	0.25000	&	0.57735	&       0.95711 &	1.37638	&       1.82735 &       2.30476 &       2.80487 & 3.32483 \\
	&$\tilde{\kappa}$&	0.24972	&	0.57726	&       0.95707 &	1.37636	&       1.82734 &       2.30476 &       2.80486 & 3.32482 \\
\hline
	&	$\Lambda$&       	&       	&               &       	&               &	-0.0984	&               &       \\
        &	$A$	&       	&       	&               &       	&               &       -.07591 &               &	\\
  IIb   &	$T$	&       	&       	&               &       	&               &       -.00333 &               &	\\
        &	$\Delta$&      		&       	&               &       	&               &$188\dotdeg357$&               &	\\
\hline
\hline
	&	$\Lambda$&		&	-5.2259	&	-5.7021	&	-5.6570	&	-4.7475	&	-0.9081	&               &       \\
        &	$A$	&       	&	.16801	&       .25000  &	.27583	&       .26477  &       .16197  &               &	\\
  IIIa	&	$T$	&       	&	.42070	&       .25000	&	.13545	&       .05269  &       -.00612 &               &	\\
	&	$\Delta$&      		&	--	&$180\dotdeg000$&	--	&$180\dotdeg000$&$187\dotdeg947$&               &	\\
        &	"	&       	&       	&               &       	&$183\dotdeg912$&       "       &               &	\\
\hline
	&	$\Lambda$ &       	&       	&               &       	&               &	-0.0623	&               &       \\
        &	$A$	&       	&       	&               &       	&               &       .08796  &               &	\\
  IIIb	&	$T$	&       	&       	&               &       	&               &       -.00358 &               &	\\
	&	$\Delta$&      		&       	&               &       	&               &$187\dotdeg575$&               &	\\
        &	"	&       	&       	&               &       	&               &       "       &               &	\\
\hline
\hline
  Type  &	$N$	&	2	&	3	&       4       &	5	&       6       &       7       &       8       & 9	\\
\hline
\end{tabular}
\vskip0.2in
Table 1. Equilibrium properties for $N$ secondaries of equal mass
sharing a circular orbit about a common primary. 
$\Lambda$ is the least non-zero eigenvalue of the frequency spectrum; 
configurations with $\Lambda < 0$ are unstable. 
$A$ is the amplitude of the star's reflex motion, 
while $T$ is the amplitude of its tidal perturbation, 
both relative to those for a single secondary 
with the mass of all $N$ secondaries combined. 
$\delta$ is the minimum angular separation between any two secondaries, 
while $\Delta$ lists all angular separations from $180^\circ$ to $190^\circ$. 
$\mu_H$ is the limiting mass ratio for stability of Type Ia equilibria 
estimated from Eq. (3). $\kappa$ is the dimensionless coefficient 
for Type IIa equilibria defined by Eq. (14), 
while $\bar{\kappa}$ is its approximation from Eq. (15). 

\newpage

Note that the stability properties for $N=7$ are particularly complicated. 
For $N$=7, solution Ia is robustly stable, with a $\Lambda$ of +9.9030,
while equilibrium IIa is just barely stable, with $\Lambda$ = +0.0426.
In contrast, solution IIIa is rather unstable, with $\Lambda$ = -0.9081,
while equilibria IIb and IIIb are only marginally unstable,
with $\Lambda$ = -0.0984 and -0.0629, respectively.

A significant perturbation of a Salo system
(such as a massive binary companion, or tides in an extended primary)
would alter the equilibrium configurations as well as their stability properties,
or could destroy some of the equilibria, or might even introduce new ones!
For example, Renner \& Sicardy (2004) have found new solutions to the problems
with $N$=3, 4, and 5 when one secondary is 100 times more massive than the others.
Furthermore, it is known that a triaxial primary in synchronous rotation
can stabilize the strongly unstable configuration IIa for $N$=2 
($\Lambda$ = -5.2500; {\it e.g.} Scheeres, 1994), 
while an oblate primary somewhat lowers the critical values of the mass ratio 
$\mu_i$ (see below) for family IIa for large values of $N$ (Vanderbei, 2008). 

Conceivably, a sufficient perturbation might completely destabilize 
the barely stable configuration IIa for $N$=7, 
or it might stabilize the marginally unstable equilibria IIb or IIIb. 
Of course, artificial station-keeping could maintain a system of secondaries 
in unstable equilibrium with relatively little effort, much as certain 
interplanetary probes (such as ISEE-3, SOHO, ACE, WMAP, and Genesis) are kept 
in orbits near the L1 and L2 Lagrange points at the edges of Earth's Hill sphere.
In the present context, however, such an arrangement would imply deliberate 
planetary engineering, and I prefer not to speculate on that possibility here. 

Although the periods of libration and growth times of instabilities 
depend on the scaled masses $\mu_i$ of the secondaries, note that 
the eigenvalues $\lambda$ are independent of $\mu_i$ to lowest order. 
Other criteria limit the stability of the equilibria for finite masses. 
For family IIa, Maxwell (1859) himself derived the criterion 
\begin{equation}
                        \mu_i \lta \Gamma / N^3
\end{equation}
in the limit of large $N$, where $\Gamma \approx$ 2.298 
(but beware that SY88 misquote $\Gamma$ as 2.23). 

For small $N$, $\Gamma$ gradually rises to $\sim$2.452 at $N$=7 
(Vanderbei \& Kolemen 2007, Table 1; but note that they use $\gamma$ 
rather than $\Gamma$, although they use $\gamma$ for $n_0$ as well. 
See also Scheeres \& Vinh 1991, 
but note that they use $\gamma$ for Euler's constant $\sim$0.5772). 
Note also that if $\mu_i$ is defined as $m_i/M$ rather than as $m_i/m_a$, 
where $M=m_a+Nm_i$ is the total mass of the system, 
the correspondingly revised $\Gamma$ is nearly independent of $N$, 
rising only to $\sim$2.335 for $N$=7. 

Eq. (2) above lends itself to interpretation in terms of the Hill sphere spacing criterion, 
a rule of thumb which states that multiple secondaries orbiting in the same plane remain stable, 
provided that they remain separated by about four or more mutual Hill radii (see Paper 1). 
The individual Hill radius of each secondary corresponding to Eq. (2) is 
$R_H$ $\equiv$ $r[\mu_i/3]^{1/3}$ $\lta$ $r \left[ \frac{\Gamma}{3 N^3} \right] ^{1/3}$ 
= $[\Gamma/3]^{1/3} r/N$, while the corresponding mutual Hill radius 
of any two secondaries is $R_\mu$ = $r[2\mu_i/3]^{1/3} \lta [2\Gamma/3]^{1/3} r/N$. 

\newpage

Meanwhile, setting $\delta=2\pi/N$ in Fig. 2, 
the distance between secondaries is just $d$ = $2r \sin (\pi/N)$ $\approx$ 
$N R_H [24/\Gamma]^{1/3} \sin(\pi/N)$ = $N R_\mu [12/\Gamma]^{1/3} \sin(\pi/N)$. 
This approaches $d \approx 2\pi r/N \approx$ 6.867 $R_H \approx$ 5.450 $R_\mu$ 
for large $N$, gradually decreasing to $2r \sin (\pi/7) \approx$ 
0.8678 $r \approx$ 6.497 $R_H \approx$ 5.157 $R_\mu$ for $N$=7 
(or to $\sim$6.603 $R_H \approx$ 5.241 $R_\mu$ as based on the total mass). 

Analogous mass-dependent stability criteria do not seem to be known 
for the Type Ia equilibria, except in the Trojan case $N$=2 (see Eq. 1); 
in terms of mutual Hill radii, Trojan systems are unstable 
for separations $\lta$ 4.3 $R_\mu$. By numerical integrations, 
SY88 found that the $N$=3 case of the Type Ia configuration became unstable 
once the minimum separation between librating secondaries became comparable 
to $\sim$5 $R_\mu$. Furthermore, SY88 also found a similar result 
for systems with three co-orbital secondaries of different masses; 
see their Fig. 15 (but note that the ``vertical line'' 
to which its caption refers is actually horizontal). 
Recently, \'{C}uk {\it et al.} (2012) determined numerically 
that test particles on horseshoe orbits also become unstable
when they approach within $\sim5$ Hill radii of the secondary 
in the Plane Circular Restricted Three-Body Problem. 

On these grounds I am embolded to apply the Hill sphere spacing criterion 
to estimate the mass stability limits for all Type Ia equilibria. 
First, I find the minimum angular separation $\delta$ between 
any two adjacent secondaries for each solution in Table 1 of SY88. 
In solution IIb for $N$=7, this $\delta \approx 45\dotdeg401$ occurs 
at longitude $\theta=180^\circ$ (see Fig. 1). For every other 
solution, the minimum separation always occurs at $\theta=0$; 
my Table 1 lists the values of $\delta$ for family Ia. 
Next, I find the corresponding minimum physical distance 
$d = 2 r \sin (\delta/2)$ for family Ia; see Fig. 2. 
Finally, I estimate that the secondaries are spaced at least five mutual 
Hill radii apart, so I set $d \gta 5 R_\mu = 5 r [2 \mu_i /3]^{1/3}$ 
and solve for the corresponding limit on $\mu_i$: 
\begin{equation}
\mu_i \lta \frac{3 d^3}{250 r^3} = \frac{12}{125} \sin^3 (\delta/2) \equiv \mu_H .
\end{equation}

Very roughly, $d \approx r \delta$ and $\mu_H \approx .012 \: \delta^3$, 
when $\delta$ is expressed in radian measure. For example, 
note that the stable equilibrium separation for $N=3$ secondaries 
of equal masses is $\sim47\dotdeg36 \approx 0.8266$ rad 
(rather than the $60^\circ$ separation of the Trojan points L4 and L5 
from a single secondary in the restricted three-body problem). 
The corresponding distance between them is $d \approx$ 0.8033 $r$ 
from Fig. 2, so $\mu_H \approx$ 0.00622 from formula (3) above. 

Table 1 also lists the values of $\mu_H$ for family Ia from Eq. (3). 
Note that $\delta$ and $\mu_H$ both decrease monotonically with increasing $N$; 
in fact, $\mu_H$ turns out to be roughly equal to $0.054/N^2$, 
scaling as the inverse square of $N$, 
rather than its inverse cube as in Maxwell's criterion (2). 
For $N=2$, my estimate $\mu_H \approx$ 0.01200 is slightly less than 
the exact value $\mu_0/2 = (1-\sqrt{8/9})/3 \approx 0.01906$ from Eq. (1); 
while for $N=3$, my estimate $\mu_H \approx$ 0.00622 
is the same as that of SY88 by construction. 

My tabulated values of $\mu_H \approx 0.00113$ for $N=7$ 
and $\mu_H \approx 0.00091$ for $N=8$ 
are several times stricter than the corresponding limits 
of $\mu_i \lta$ .00715 ($d \approx 0.8678 \, r \approx 5.156 \, R_\mu$) 
and $\mu_i \lta$ .00471 ($d \approx 0.7654 \, r \approx 5.227 \, R_\mu$) 
for family IIa, from Table 1 of Vanderbei \& Kolemen (2007); 
but this is not surprising, because the secondaries in family Ia 
are packed nearly twice as close together as those in family IIa. 
Further analytical or numerical work could clarify these relations, 
but lies beyond the scope of this paper. 

\begin{figure}
\centerline{\psfig{figure=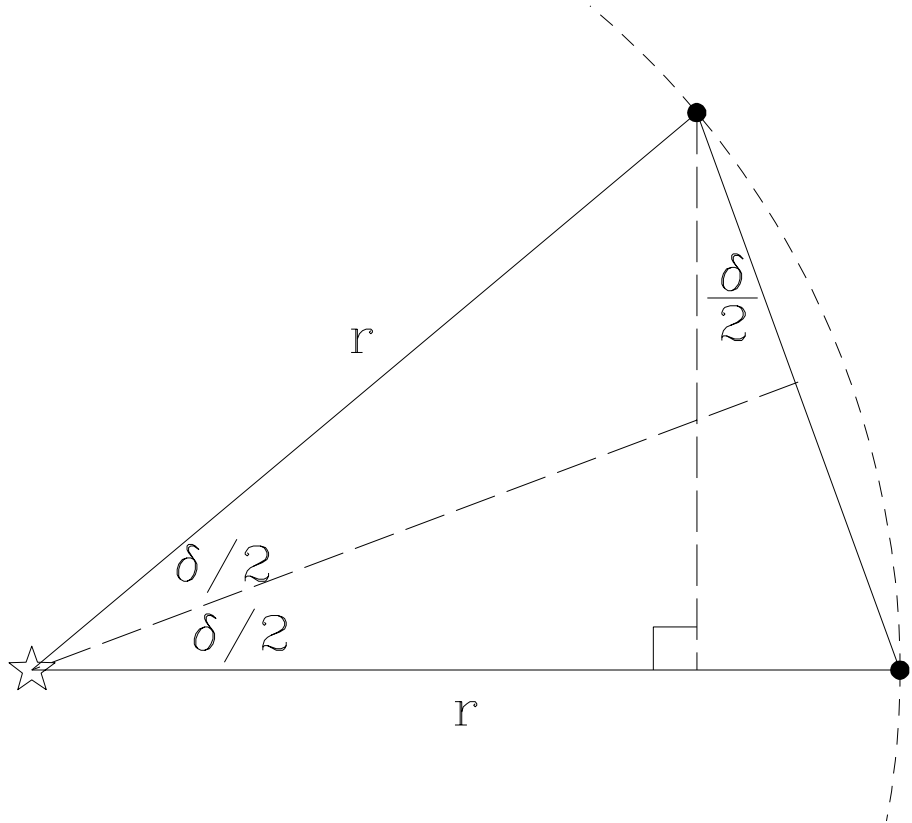,width=6.5in,height=6.5in}}
\caption{ Schematic for distance and force between two planets
at equal distances $r$ from their parent star,
but separated by a central angle $\delta$. }
\end{figure}

\newpage

\subsection{Reflex motion}

This section examines the amplitude and phase of the reflex motion for Salo systems. 
The orbital frequency $n_0$ of co-orbital systems presents certain subtleties, 
and is discussed in the appendix; but the equivalent of formula (2) still applies, 
at least to order $\mu_i \equiv m_i/m_a$. 

2.2.1: Amplitudes

For systems with multiple secondaries 
of masses $m_i$ located at positions $r_i$, 
the system center of mass (barycenter) is located at 
\begin{equation}
                {\bf A} = \sum_{i=1}^N {\bf r_i} m_i /M 
\end{equation}
relative to the center of the primary $m_a$, 
where $M \equiv m_a +\sum m_i$ is the total mass of the system. 
Note that formula (4) above is valid even if the $m_i$ and $r_i$ are unequal. 

For co-orbital systems, however, all of the secondaries 
lie at the same distance $r$ from the primary, to order $\mu_i$. 
Furthermore, for Salo systems, all of the $m_i$ are set strictly equal, 
by definition. Then Eq. (4) simplifies to 
\begin{equation}
		A = \frac{r m_i}{M} \sum_{i=1}^N \cos \theta_i
\end{equation}
(to order $\mu_i$), where $\theta_i$ is the angular location 
of $m_i$ relative to the origin of longitude (see Fig. 1). 
The direction of ${\bf A}$ lies along the axis of symmetry 
$\theta=0$ (or $\theta=\pm180^\circ$). 

Table 1 lists $A$ from formula (5) above, 
but normalized by $N r m_i/M$ rather than just $r m_i/M$, 
in order to account for the total mass $N m_i$ of secondaries. 
This table shows that Salo systems all have reduced reflex amplitudes 
$A$ (compared to unity for a single secondary of mass $N m_i$). 
Note that the equilateral triangle configuration Type Ia with 
$N$=2 equal-mass planets has the largest $A = \sqrt{3/4} \approx$ 
0.8660, only $\sim$13 percent less than that for a single planet. 

\begin{figure}
\centerline{\psfig{figure=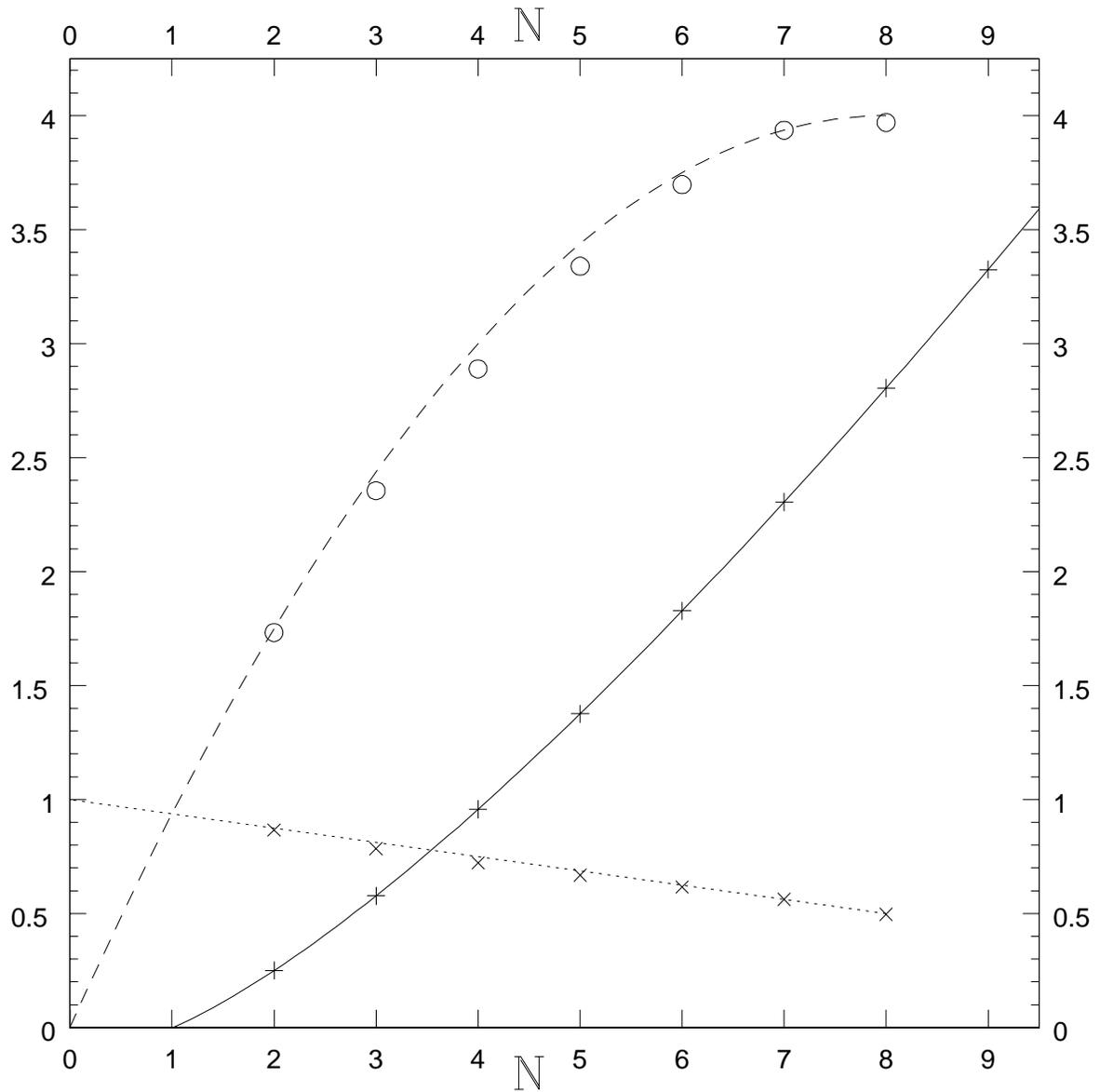,width=6.5in,height=6.5in}}
\caption{ X symbols: reflex amplitude $A$ as a function of $N$
for Type Ia solutions. Dotted line: approximation $A=1-N/16$.
Open dots: $NA$ for Type Ia solutions.
Dashed parabola: approximation $NA=N-N^2/16$.
Plus signs: $\kappa$ as a function of $N$ for Type IIa solutions.
Solid curve: approximation (15). }
\end{figure}

Figure 3 plots both $A$ and $NA$ as a function of $N$ for family Ia. 
The X's plot $A$, while the dotted line 
graphs my heuristic approximation $A \approx 1 -N/16$. The O's plot $NA$, 
while the dashed parabola graphs my approximation $NA \approx N -N^2/16$.
Note how $NA$ grows monotonically with $N$, 
but $A$ itself decreases almost linearly with increasing $N$. 
For $N$=7 or 8, for example, these systems would show almost four times 
the radial velocity signal as for an individual planet of mass $m_i$, 
but only about half of the signal of a single body 
with the combined mass $N m_i$ of all the planets. 

For comparison, note the nearly linear progression $A$= .16197, .08796, 
.00000, -.07591 for solutions IIIa, IIIb, IIa, IIb respectively with $N$=7 
(see Fig. 1). The slightly asymmetrical Type IIb solution 
has the smallest non-zero $|A|$ of all, $\sim$13 times less 
than the corresponding amplitude of unity for a single planet. 
This value of $A \approx$ -.07591 is also the only formally 
negative reflex amplitude, because the angular locations $\theta_i$ 
are slightly shifted towards $\theta=\pm180^\circ$, while all of the 
Type I and Type III configurations are shifted toward $\theta$=0. 

In contrast, Table 1 shows that the strictly symmetrical solutions 
of family IIa cause no reflex motion at all ($A$=0). 
Note that the Type IIa equilibria all possess $N$ axes of symmetry, 
while each of the other solutions has only one. 
This symmetry of the Type IIa systems nullifies 
their radial velocity and astrometric signatures (Smith \& Lissauer 2010). 

\newpage

2.2.2 Tides 

As explained in Paper 1, planets also raise tides on their parent star 
which can cause observable modulation of its brightness, 
called ``ellipsoidal variations''. From Paper 1, 
the amplitude $T$ of the tide-raising potential in a Salo system 
is simply analogous to formula (5) for the reflex motion: 
\begin{equation}
                T = \frac{r m_i}{M} \sum_{i=1}^N \cos (2 \theta_i)
\end{equation}
(to order $\mu_i$); note the doubling of the angle $\theta$. 
Table 1 also lists the tidal amplitude $T$ from formula (6) above, 
again normalized by $r N m_i/M$ rather than by $r m_i/M$, 
to account for the total mass $N m_i$ of secondaries.

Like the reflex motion $A$, symmetry implies that the tidal amplitudes $T$ 
also vanish for family IIa, except for the $N$=2 member, which consists 
of a pair of secondaries on diametrically opposite sides of the primary 
(located at $\theta_i$=0 and $180^\circ$ for consistency with family III). 
For this particular configuration, the net tidal amplitude ($T$=1) 
is the same as that of a single companion with the combined mass 
of both secondaries, but twice that of either secondary alone ($NT$=2). 
If such a system were stabilized somehow, 
it would show its full tidal amplitude $T$, but no reflex amplitude $A$. 

In comparison, all of the other Salo systems have reduced tidal amplitudes. 
For example, the Trojan configuration Ia for $N$=2, 
$m_i=m_b=m_c$ has a tidal amplitude $T$=1/2. 
Thus each secondary has only half of its usual tidal effect, 
but both of them together raise a total tide $NT$=1 
as great as that due to a single body of mass $m_i$. 
All of the remaining tidal amplitudes are even smaller. 

For families Ia and IIIa, $T$ and $NT$ both decline monotonically 
with increasing $N$ (in the algebraic sense for family Ia, and also in 
the absolute sense for family IIIa). In fact, all of their tidal amplitudes 
become formally negative for larger $N$. For $T$ and $A$ of the same sign, 
the maximum tidal perturbations are aligned with the radial displacement; 
but for $T$ and $A$ of opposite signs, 
the tidal perturbations are actually orthogonal to the displacement. 
The latter mimics the effect of a single secondary with negative mass! 

Negative tidal amplitudes actually arise when more secondaries lie 
near $\theta=\pm90^\circ$ than lie near $\theta$=0 or $180^\circ$. 
Thus configuration Ia with $N$=8 has the most negative $T$ and $NT$, 
at -0.10231 and -0.81851, respectively. For comparison, the slightly 
asymmetrical Type IIb solution with $N$=7 also has the smallest non-
zero $|T| \approx 0.00333$ and $|NT| \approx 0.0233$, $\sim$43 times 
less than the corresponding amplitude $NT$=1 for a single secondary. 
Among the stable configurations, the lop-sided Type Ia solution with $N$=6 
has the smallest non-zero $|T| \approx 0.01431$ and $|NT| \approx 0.0859$, 
$\sim$12 times less than the corresponding amplitude for a single secondary. 

\newpage

2.2.3: Phases

As Fig. 1 shows, the Type Ia configuration for seven planets 
spreads over slightly more than a semicircle of their mutual orbit; 
the angular separation between the first and last planets is 
$180\dotdeg552$ (or $179\dotdeg448 = 360^\circ -180\dotdeg552$). 
The closeness of this interval to $180^\circ$ could complicate 
the interpretation of transits and occultations. For example, 
the first planet would pass through transit only half a degree 
of orbital phase before the last planet underwent occultation. 
That occultation might be mistaken for the shallow transit 
of a satellite, for instance. 

To explore such possibilities further, I define $\Delta$ as 
the angular interval between any two planets in the same orbit. 
Table 1 lists all such intervals between $180^\circ$ and $190^\circ$ 
inclusive occurring in all of the equilibria, except Type IIa. 
The dashes in the table (--) signify that the specified configuration 
contains no such interval, while the ditto marks (") indicate that 
the specified configuration contains two separate but equal intervals. 

Note that solution IIIa for $N$=6 contains one interval 
of $183\dotdeg912$, and another of exactly $180^\circ$, 
corresponding to a pair of diametrically opposite secondaries. 
In the latter case, transits and occultations 
would be simultaneous (in the absence of librations). 
Configuration IIIa for $N$=4 also has one interval of 
exactly $180^\circ$, while every Type IIa equilibrium contains 
$N/2$ such intervals if $N$ is even, but none if $N$ is odd. 

\section{CONCLUSIONS}

Radial velocity measurements can seriously misjudge 
the masses of co-orbital exoplanets, or in some cases 
fail to detect them altogether. In Salo systems, 
for example, where multiple planets of equal masses 
share a circular orbit about the same star, 
certain stable configurations produce radial velocity variations 
only half as great as a single planet of the same total mass would, 
but almost four times greater than each individual planet does. 

In comparison, tides raised by these planets can cause ellipsoidal 
stellar brightness variations about 12 times as great as a single planet 
of the same mass would, and of opposite sign as well! 
Other stable configurations produce no radial velocity signal 
or ellipsoidal variations at all, and could be overlooked entirely, 
unless they happen to be detected by the transit method. The transit 
method itself also can produce misleading results in Salo systems. 

Of course it is not likely that co-orbital planets all would have the same mass, 
but the problems for Salo systems are representative of those for a wide variety 
of more natural systems. Most algorithms currently in use to analyze radial velocity 
and transit data assume that no two planets share the same orbital period. 
Thus co-orbital exoplanets can effectively hide from detection, 
or produce misleading conclusions. 

\newpage
		Appendix: Orbital Frequency

The orbital angular frequency of a co-orbital system, 
{\it a.k.a.} its mean motion $n_0$, 
is related to the centripetal force ${\bf F}_a$ 
on the primary $m_a$ toward the system barycenter: 
\begin{equation}
		{\bf F}_a = m_a {\bf A} n_0^2 ,
\end{equation}
where ${\bf A}$ is the location of the barycenter 
relative to the primary from Eq. (4) of the main text. 
The source of this force is the gravitational attraction of all the secondaries $m_i$: 
\begin{equation}
{\bf F}_a = \sum_{i=1}^N G m_a m_i {\bf r_i} /r_i^3 \approx G m_a M {\bf A} /r^3 ,
\end{equation}
since each of the $r_i$ is equal to $r$ (to order $\mu_i \equiv m_i/m_a$). 
Then equating the gravitational force (8) above to the centripetal force (7), 
dividing both sides by $m_a$, and cancelling ${\bf A}$, leaves 
\begin{equation}
                n_0^2 \approx G M /r^3 = G [ m_a +\sum m_i ] /r^3
\end{equation}
to order $\mu_i$. 

Note that result (9) above applies 
even if the secondary masses $m_i$ are not all equal, as long as $A \ne 0$. 
There is at least one situation in which the above argument fails, though: 
For the Type IIa Salo configurations, with $N$ secondaries 
of equal masses $m_i$, $\bf A$ vanishes by symmetry, 
and cannot be cancelled from Eqs. (7) and (8). 
In this circumstance we may consider the centripetal force ${\bf F}_N$ 
on one of the $N$ secondaries instead: 
\begin{equation}
                {\bf F}_N = -m_i {\bf r}_N n_0^2 ,
\end{equation}
where we have chosen the index $i=N$ without loss of generality. 

This centripetal force is equal to the net gravitational force on $m_N$ 
from the central primary as well as from all of the other $N$-1 secondaries: 
\begin{equation}
                        {\bf F}_N = -G m_a m_i {\bf r}_N /r^3
                +\sum_{j=1}^{N-1} G m_i^2 ({\bf r}_j - {\bf r}_N) / d_j^3 ,
\end{equation}
where $d_j \equiv |{\bf r}_j - {\bf r}_N|$ is the distance 
between $m_j$ and $m_N$. By geometry (see Fig. 2), we have 
$d_j = 2 r \sin (\delta/2)$, where $\delta = 360^\circ \times j/N = 2 \pi j/N$ 
radians is the angular separation between $m_j$ and $m_N$. 
Then considering only the components of ${\bf F}_N$ along ${\bf r}_N$ 
by symmetry (see Fig. 2 again), we find 
\begin{equation}
                        {\bf F}_N = -G m_a m_i {\bf r}_N /r^3
        -\sum_{j=1}^{N-1} \frac{ G m_i^2 {\bf r}_N }{ 4 r^3 \sin (\pi j/N) } .
\end{equation}

Now equating expressions (10) and (12) above, 
and cancelling $-m_i {\bf r}_N$, leaves 
\begin{equation}
                        n_0^2 = G[m_a +\kappa m_i]/r^3 ,
\end{equation}
                where the dimensionless coefficient $\kappa$ is given by 
\begin{equation}
                \kappa = \frac{1}{4} \sum_{j=1}^{N-1} \csc(\pi j/N) .
\end{equation}
For example, $\kappa$ = 1/4 for $N$=2, 
$\sqrt{1/3}$ for $N$=3, $1/4 +\sqrt{1/2}$ for $N$=4, and so on. 
The numerical values of $\kappa$ for $2 \le N \le 9$ are listed 
in the middle row of Table 1, and plotted as the plus signs in Fig. 3. 
As defined above, $\kappa$ is equal to $K/4$ of Scheeres \& Vinh (1991) 
or to $S/4$ of Scheeres \& Vinh (1993), 
and identical to $I_N$ of Vanderbei \& Kolemen (2007) and of Vanderbei (2008). 

For large $N$, Scheeres \& Vinh (1991, Eq. A.17) find the analytic approximation 
\begin{equation}
	\kappa \approx \frac{N}{2\pi} \left[ \ln \left( \frac{2 N}{\pi} \right) 
		+\gamma \right] -\frac{\pi}{144 N} \equiv \tilde{\kappa} 
\end{equation}
plus terms of order $1/N^3$ or higher, 
where $\gamma \approx$ 0.5772 is widely known as Euler's constant. 
I have verified formula (15) above numerically for $2 \le N \le 1000$ 
and $N \in$ $\{ 10^4, 10^5, 10^6, 10^7, 10^8, 10^9 \}$; 
this approximation turns out to be quite accurate even for small $N$. 
To demonstrate, the results of formula (15) above are listed 
in Table 1 as well, and plotted as the solid curve in Fig. 3. 
The simpler approximation $\kappa \approx \frac{N \ln N}{2\pi}$ 
also is passable, but Vanderbei \& Kolemen (2007, Eq. 22) 
give the much less accurate $I_N \approx \frac{N \ln (N/2)}{2\pi}$. 
(Also note that every $\alpha$ should be squared 
in Eq. 39 of Vanderbei \& Kolemen, 2007.) 

Comparing formula (13) for family IIa with approximation (9) 
for asymmetric co-orbital configurations shows that $\kappa m_i$ 
takes the place of the total mass of secondaries $\sum m_i$ in the 
mean-motion formula, or of $N m_i$ for Salo systems with all $m_i$ equal. 
It turns out that $\kappa < N$ for $N \le 472$, so that Eq. (9) overestimates 
$n_0$; but $\kappa > N$ for $N \ge 473$, so that Eq. (9) underestimates $n_0$. 

\newpage

\begin{center}
				REFERENCES
\end{center}

\setlength{\parindent}{-0.3in}

\'{C}uk, M., D. P. Hamilton, and M. H. Holman, 2012. 
        Long-term stability of horseshoe orbits. 
        {\it M.N.R.A.S.} {\bf 426}, 3051--3056. 

Dobrovolskis, A. R., 2013 (Paper 1). Effects of Trojan exoplanets 
	\\ on the reflex motions of their parent stars. 
	{\it Icarus} {\bf 226}, 1635--1641. 

Lagrange, J. L., 1772. Essai sur le probl\`{e}me des trois corps. \\ 
	{\it Prix de l'Academie des Sciences de Paris} {\bf 9}, 229--331. 

Maxwell, J. C., 1859. On the stability of motions of Saturn's rings. 
	MacMillan, Cambridge, UK. 

Renner, S., and B. Sicardy, 2004. Stationary configurations 
	for co-orbital satellites \\ with small arbitrary masses. 
	{\it Cel. Mech. \& Dyn. Astron.} {\bf 88}, 397--414. 

Routh, E. J., 1875. On Laplace's three particles, \\ 
	with a supplement on the stability of steady motion. 
	{\it Proc. London Math. Soc.} {\bf 6}, 86--97. 

Salo, H., and C. F. Yoder, 1988 (SY88). 
	The dynamics of coorbital satellite systems. \\ 
	{\it Astronomy \& Astrophysics} {\bf 205}, 309--327. 

Scheeres, D. J., and N. X. Vinh, 1991. 
	Linear stability of a self-gravitating ring. \\ 
	{\it Cel. Mech. \& Dyn. Astron.} {\bf 51}, 83--103. 

Scheeres, D. J., and N. X. Vinh, 1993. 
	The restricted $P$ + 2 body problem. \\ 
	{\it Acta Astronautica} {\bf 29}, 237--248. 

Scheeres, D. J., 1994. Dynamics about uniformly rotating triaxial ellipsoids: 
	\\ Applications to asteroids. {\it Icarus} {\bf 110}, 225--238. 

Smith, A. W., and J. J. Lissauer, 2010. 
	Orbital stability of systems of closely-spaced planets, II: 
	configurations with coorbital planets. 
	{\it Cel. Mech. \& Dyn. Astron.} {\bf 107}, 487--500. 

Vanderbei, R. J., and E. Kolemen, 2007. Linear stability of ring systems. 
        \\ {\it Astron. J.} {\bf 133}, 656--664. 

Vanderbei, R. J., 2008. 
	Linear stability of ring systems about oblate central masses. 
	\\ {\it Advances in Space Research} {\bf 42}, 1370--1377. 

\end{document}